\documentstyle[prb,aps,psfig,twocolumn,preprint]{revtex}
\begin{document}
%\draft

\onecolumn

\title{Analytical Hartree-Fock gradients for periodic systems}
\author{K. Doll$^1$, V. R. Saunders$^1$ and N. M. Harrison$^{1,2}$ } 
\address{$^1$CLRC, Daresbury Laboratory, Daresbury, Warrington, WA4 4AD, UK}
\address{$^2$ Department of Chemistry, Imperial College, London, SW7 2AY, UK} 

\maketitle

\begin{abstract}
We present the theory of analytical Hartree-Fock gradients for periodic
systems as implemented in the code CRYSTAL.
We demonstrate how derivatives
of the integrals can be computed with the McMurchie-Davidson algorithm.
Highly accurate gradients with respect to nuclear coordinates 
are obtained for systems periodic in 0,1,2 or 3
dimensions.
\end{abstract}

\pacs{ }

\narrowtext
\section{Introduction}

The determination of equilibrium structure is one of the most important
targets in electronic structure calculations. Analytical gradients
provide an important tool to facilitate this and therefore the
implementation of analytical gradients has become an important part
of modern codes. 
Although most solid state calculations are nowadays performed within
the framework of density functional theory, Hartree-Fock theory
can serve as a useful starting point for a correlation treatment.
In the field of quantum chemistry, 
a Hartree-Fock solution is necessary to make a wavefunction based 
correlation scheme such as, for example, the coupled-cluster approach,
applicable. Therefore, the determination of a Hartree-Fock solution is
often an important target. 

The calculation of Hartree-Fock gradients
was pioneered by Pulay who performed the first implementation for
multicenter basis sets \cite{Pulay}. It should be mentioned,
that the theory had already been derived earlier independently \cite{Bratoz}.
Analytical gradients have become
an important area in quantum chemistry and several review articles
have been published \cite{PulayAdv,PulayChapter,Schlegel,Helgaker}.

Significant work has already been performed for one-dimensional systems:
formulas for analytic gradients,  
with respect to nuclear 
coordinates as well as with respect to the lattice vector, have been derived
and implemented in a periodic code \cite{Teramae};
and the theory has been extended to metallic 
systems\cite{Kertesz}. Further progress has been the derivation and
implementation of formulas for MP2 energy  \cite{Suhai,SunBartlett} 
and  gradients
\cite{HirataIwata}, as well as gradients on the density functional
level \cite{HirataIwataDFT}. Even formulas for
second derivatives have meanwhile been coded
\cite{HirataIwata2nd}. Recently, a scheme for an
 accurate treatment of long-range Coulombic
effects in Hartree-Fock gradients has been presented
\cite{Jacquemin}, and a new implementation of density functional
energy and gradients for periodic systems has been 
demonstated to be highly efficient and accurate \cite{KudinScuseria}.

In this article, we report on an implementation of
Hartree-Fock gradients with respect to nuclear coordinates
in a general periodic code (the 
CRYSTAL\cite{Manual,Pisani1980,CRYSTALbuch,Vicbook}
package), which is to the best of 
our knowledge the first implementation for the
case of 2- and 3- dimensional periodicity.

The article is structured as follows: in section \ref{BasisHFsection},
the basis functions and Hartree-Fock equations are given.
The calculation of integrals
which relies on the McMurchie-Davidson algorithm
and the calculation of gradients of the integrals
is explained in sections \ref{Integralsection} and 
\ref{Calculationofderivatives}. 
The total energy as calculated by the CRYSTAL code is given in
section \ref{FMexplizit}.
In section \ref{Gradienttotenysection}, we explain the calculation 
of forces and possible sources of error. Finally, in 
section \ref{applications} we illustrate the accuracy of the gradients with 
some examples.

\section{Basis function and Hartree-Fock equations}

\label{BasisHFsection}

In this section, we  summarize the basis functions used in
the CRYSTAL code and give the structure of the Hartree-Fock equations.

\subsection{Basis functions}

Unnormalized spherical Gaussian type functions (SGTF)
in a polar coordinate system 
characterized by the set of variables $(|\vec r|,\vartheta,\varphi)$,
and centered at $\vec A$,
are defined as

\begin{equation}
S(\alpha,\vec r-\vec A,n,l,m)={|\vec r-\vec A|}^{2n+l} 
 {\rm P}_l^{|m|}(\cos \ \vartheta)
\exp({\rm i}m\varphi)\exp (-\alpha {|\vec r-\vec A|}^2)
\end{equation}

with ${\rm P}_l^{|m|}$ being the associated Legendre function.
In the context of the McMurchie-Davidson algorithm, 
Hermite Gaussian type functions are necessary which are defined as:

\begin{eqnarray}
\Lambda(\gamma,\vec r-\vec A,t,u,v)=
\bigg(\frac{\partial}{\partial A_x}\bigg)^t
\bigg(\frac{\partial}{\partial A_y}\bigg)^u
\bigg(\frac{\partial}{\partial A_z}\bigg)^v {\exp}(-\gamma
|\vec r-\vec A|^2)
\end{eqnarray}

Real spherical Gaussian type functions are
defined as

\begin{eqnarray*}
R(\alpha,\vec r-\vec A,n,l,0)=S(\alpha,\vec r-\vec A,n,l,0)\\
R(\alpha,\vec r-\vec A,n,l,|m|)={\rm Re \ } S(\alpha,\vec r-\vec A,n,l,|m|)\\
R(\alpha,\vec r-\vec A,n,l,-|m|)={\rm Im \ } S(\alpha,\vec r-\vec A,n,l,|m|)
\end{eqnarray*}

CRYSTAL uses real spherical Gaussian type functions,
which are in the following denoted as $\phi_{\mu}(\vec  r - \vec A_{\mu})=
N_{\mu} R(\alpha,\vec A_{\mu},n,l,m)$, 
with the normalization $N_{\mu}$. $\mu$ is an index enumerating
the basis functions
in the reference cell (e.g. the primitive unit cell).
Although the code allows only the use
of SGTFs with $n=0$,
in the process of the
evaluation of molecular integrals, 
SGTFs with $n \ne 0$ are used \cite{VicNATO}.

\subsection{Hartree-Fock equations}

The Hartree-Fock treatment of periodic systems \cite{Andre} is briefly 
repeated in this section.
We assume that orbitals are doubly occupied and work within
the restricted Hartree-Fock formalism.
The crystalline orbitals are linear combinations of Bloch functions

\begin{equation}
\Psi_i(\vec r, \vec k)=\sum_{\mu} a_{\mu i}(\vec k)\psi_{\mu}(\vec r, \vec k)
\end{equation}

which are expanded in terms of real spherical Gaussian type functions
with fixed contraction coefficients $d_j$:

\begin{equation}
\psi_{\mu}(\vec r, \vec k)=N_{\mu} \sum_{\vec g} \sum_j d_j
R(\alpha_j,\vec r-\vec A_{\mu}-\vec g,n,l,m)
{\rm e}^{{\rm i} \vec k\vec g}
\end{equation}

The sum over $\vec g$ is over all direct lattice vectors.
The Hartree-Fock-Roothaan 
equations have a structure similar to the molecular
case and have to be solved on a set of points in the reciprocal lattice:

\begin{eqnarray}
\label{HFequation}
\sum_{\nu}F_{\mu \nu}(\vec k)a_{\nu i}(\vec k)=\sum_{\nu}
S_{\mu \nu}(\vec k)a_{\nu i}(\vec k)\epsilon_i(\vec k)
\end{eqnarray}

The Fock matrix $F$ is given in detail in section
\ref{FMexplizit} of this article, the overlap matrix $S$ in section
\ref{Integralsection}.
The spin-free density matrix in reciprocal space is defined as

\begin{eqnarray}
P_{\mu\nu}(\vec k)=2\sum_{i}a_{\mu i}(\vec k)a_{\nu i}^*(\vec k)
\Theta(\epsilon_F-\epsilon_i(\vec k))
\end{eqnarray}

with the Fermi energy $\epsilon_F$ and the Heaviside function $\Theta$; $i$
is an index enumerating the eigenvalues.
The density matrix in real space 
$P_{\mu\vec 0\nu\vec g}$
is obtained by Fourier transformation.

\section{Calculation of integrals}

\label{Integralsection}

\subsection{Types of integrals}

In this section we  summarize the appearing types of integrals.
This is done with the assumption that the basis functions, $\phi_{\mu}$, are
real.

\subsubsection{Overlap integral}

The basic integral is the overlap integral:

\begin{equation}
S_{\mu\vec {g_1}\nu\vec {g_2}}= 
\int \phi_{\mu}(\vec r - \vec A_{\mu}-\vec g_1)
\phi_{\nu}(\vec r - \vec A_{\nu}-\vec g_2){\rm d^3r}
\end{equation}

 The integration is over the whole space, i.e. $x$ from $-\infty$ to $+\infty$
and similarly for $y$ and $z$.
Exploiting translational invariance we can rewrite this as

\begin{equation}
S_{\mu\vec {0}\nu\vec {g}}=
\int \phi_{\mu}(\vec r - \vec A_{\mu})
\phi_{\nu}(\vec r - \vec A_{\nu}-\vec g){\rm d^3r}
\end{equation}

with $\vec g=\vec {g_2}-\vec {g_1}$. 

Further integrals appearing are:

\subsubsection{Kinetic energy integrals}

\begin{equation}
T_{\mu\vec 0\nu\vec g}= \int \phi_{\mu}(\vec r-\vec A_{\mu}
)(-\frac{1}{2}\Delta_{\vec r})\phi_{\nu}(\vec r - \vec A_{\nu}-\vec g)
{\rm d^3r}
\end{equation}

\subsubsection{Nuclear attraction integrals}

\begin{equation}
N_{\mu\vec 0\nu\vec g}= -\sum_a Z_a\int \phi_{\mu}(\vec r-\vec A_{\mu})
A(\vec r-\vec A_{a})
\phi_{\nu}(\vec r - \vec A_{\nu}-\vec g) {\rm d^3r}
\end{equation}

where $A$ is defined as the Coulomb potential in the molecular case,
as the Euler-MacLaurin potential for systems periodic in one dimension
\cite{Delhalle,Vic1994}, as Parry's potential \cite{Parry} for
systems periodic in two dimensions, and as the Ewald potential for
systems periodic in three dimensions \cite{Ewald,VicCoulomb}.
The $a$ summation runs over all nuclei of the primitive unit cell.

\subsubsection{Electron-electron Coulomb interaction integrals}

A bielectronic integral can be defined as

\begin{equation}
B_{\mu\vec 0\nu\vec g\tau\vec n \sigma\vec n+\vec h}=
\int\frac {\phi_{\mu}(\vec r-\vec A_{\mu})\phi_{\nu}(\vec r-\vec A_{\nu}-
\vec g)
\phi_{\tau}(\vec r \: '-\vec A_{\tau}-\vec n
)\phi_{\sigma}(\vec r \: ' - \vec A_{\sigma}-\vec n-\vec h)}
{|\vec r-\vec r \: '|} {\rm d^3r \: d^3r'}
\end{equation}

In the context of periodic systems, it is necessary to perform summation
over all lattice vectors $\vec g, \vec h, \vec n$. We define a Coulomb
integral as follows

\begin{eqnarray} & &
C_{\mu\vec 0\nu\vec g\tau\vec 0\sigma \vec h}
= \sum_{\vec n}^{pen}
\int\frac {\phi_{\mu}(\vec r-\vec A_{\mu})\phi_{\nu}(\vec r-\vec A_{\nu}-
\vec g)
\phi_{\tau}(\vec r \: '-\vec A_{\tau}-\vec n
)\phi_{\sigma}(\vec r \: ' - \vec A_{\sigma}-\vec n-\vec h)}
{|\vec r-\vec r \: '|}{\rm d^3r \: d^3r'}= \nonumber \\ & &
\sum_{\vec n}^{pen}B_{\mu\vec 0\nu\vec g\tau\vec n \sigma\vec n+\vec h}
\end{eqnarray}

The penetration depth $pen$ is defined as those terms for which

\begin{equation}
\label{T2equation}
\int[g(\alpha_{\mu}^{min},\vec r-\vec A_{\mu})
g(\alpha_{\nu}^{min},\vec r-\vec A_{\nu}-\vec g)]
^{1/2}g(\alpha_{\tau}^{min},\vec r-\vec A_{\tau}-\vec n) {\rm d^3r} > 
10^{-ITOL2}
\end{equation}

holds, with
$g(\alpha_a^{min},\vec r)=(\frac{2\alpha_a^{min}}{\pi})^{3/4}{\exp}
({-\alpha_a^{min} |\vec r|^2})$.

$g(\alpha_a^{min},\vec r)$ means the lowest exponent of all
Gaussians centered at $\vec A_a$.
For these integrals, the Coulomb interaction is evaluated
without approximation. All the other integrals are evaluated 
with a multipolar expansion.
"$ITOL2$" is a tolerance which can be chosen by the user of the code.
This criterion introduces an asymmetry in the energy expression: 
a given bielectronic integral
might be evaluated in different ways
for $B_{\mu\vec 0\nu\vec g\tau\vec n \sigma\vec n+ \vec h}$ and
$B_{\tau\vec n\sigma \vec n+\vec h \mu\vec 0\nu\vec g}$. 
Avoiding this would however be
very inefficient: To keep the symmetry, the lattice sum would have
to be further broken down into pieces which are evaluated exactly and other
pieces which are approximated --- this would require a much higher 
computational effort and more disk storage. The simpler criterion in
equation \ref{T2equation} minimizes the effort, and, when $ITOL2$ is
chosen sufficiently large, the violation of the symmetry is 
negligible. We will illustrate this with some examples in section
\ref{applications}.

\subsubsection{Electron-electron exchange interaction integrals}

\begin{eqnarray}
& &
X_{\mu\vec 0\nu\vec g\tau\vec 0\sigma \vec h}
= \sum_{\vec n}
\int\frac {\phi_{\mu}(\vec r-\vec A_{\mu})
\phi_{\tau}(\vec r-\vec A_{\tau}-\vec n)
\phi_{\nu}(\vec r \: '-\vec A_{\nu}-\vec g)
\phi_{\sigma}(\vec r \: ' - \vec A_{\sigma}-\vec n-\vec h)
}
{|\vec r-\vec r \: '|}{\rm d^3r d^3r'}=\nonumber \\  & &
\sum_{\vec n}
B_{\mu\vec 0\tau\vec n \nu\vec g\sigma\vec n+\vec h} 
\end{eqnarray}

For an individual exchange integral, 

\begin{eqnarray}
\label{Exchangeequality}
B_{\mu\vec 0\tau\vec n\nu\vec g\sigma\vec n+\vec h} =
B_{\tau\vec n\mu\vec 0 \sigma \vec n+\vec h\nu\vec g} 
\end{eqnarray}

should hold. 
However, for efficiency reasons, two different thresholds have been
introduced \cite{Manual} which leads to another possible asymmetry:
an exchange integral is discarded when the pseudooverlap associated with
$\phi_{\mu}(\vec r-\vec A_{\mu})$ and $\phi_{\nu}(\vec r-\vec A_{\nu}-\vec g)$
or the pseudooverlap associated with
$\phi_{\tau}(\vec r-\vec A_{\tau}-\vec n)$ and
$\phi_{\sigma}(\vec r - \vec A_{\sigma}-\vec n-\vec h)$ is
smaller than certain thresholds, $10^{-ITOL4}$ and $10^{-ITOL5}$. 
It is recommended that the threshold $ITOL5$ associated with 
$\phi_{\tau}(\vec r-\vec A_{\tau}-\vec n)$ and
$\phi_{\sigma}(\vec r - \vec A_{\sigma}-\vec n-\vec h)$
should be higher than $ITOL4$. This, however, will lead to a violation
of equation \ref{Exchangeequality} and therefore another asymmetry in
the energy expression. A further cutoff parameter, $ITOL3$, selects the
exchange integrals symmetrically: exchange integrals are also neglected if the
overlap of $\phi_{\mu}(\vec r-\vec A_{\mu})$ with
$\phi_{\tau}(\vec r-\vec A_{\tau}-\vec n)$ or the overlap of
$\phi_{\nu}(\vec r-\vec A_{\nu}-\vec g)$ with
$\phi_{\sigma}(\vec r - \vec A_{\sigma}-\vec n-\vec h)$ 
is lower as $10^{-ITOL3}$. 
This is a symmetric cutoff, and therefore should
not lead to an inaccuracy in the forces.

\subsubsection{Multipolar integrals}

The charge distribution is approximated with the help of a multipolar
expansion up to order $L$.

\begin{equation}
\eta_l^m(\rho;\vec A_c)=\int \rho(\vec r) X_l^m(\vec r-\vec A_{c}){\rm d^3r}
\end{equation}

with $X_l^m$ being regular solid harmonics \cite{VicCoulomb} and
the charge $\rho(\vec r)$ defined as

\begin{equation}
\rho(\vec r)=-\sum_{\vec g,\mu,\nu}P_{\nu\vec g\mu\vec 0}
\phi_{\mu}(\vec r-\vec A_{\mu})\phi_{\nu}(\vec r-\vec A_\nu-\vec g)
\end{equation}

\subsubsection{Field integrals}

The electrostatic potential is approximated with an expansion up to
the maximum quantum number $L$.

\begin{equation}
M_{l\mu\vec 0\nu\vec gc}^m=Z_l^m(\hat{\vec A_c})\int
\phi_{\mu}(\vec r-\vec A_{\mu})\phi_{\nu}(\vec r-\vec A_\nu-\vec g)
\bigg[A(\vec r-\vec A_c)-\sum_{\vec n}^{pen}
\frac{1}{|\vec r-\vec A_c-\vec n|}\bigg]{\rm d^3r}
\end{equation}

with $Z_l^m(\hat {\vec A_c})$ being the spherical gradient operator,
$Z_l^m(\hat {\vec A_c})=\frac{N_l^m}{(2l-1)!!}X_l^m(\hat {\vec A_c})$ 
and a normalization $N_l^m$ (Ref. \onlinecite{VicCoulomb}).

\subsubsection{Spheropole}

This term arises because the charge distribution is approximated by
a model charge distribution in the long range. However, the
use of the Ewald potential instead of the Coulomb potential requires
a correction in the three-dimensional case\cite{VicCoulomb}.

\begin{equation}
Q=\sum_{c}Q_c=
\sum_{c}\frac{2\pi}{3V}\int(\rho_c(\vec r)-\rho_c^{\rm model}(\vec r))
|\vec r|^2 
{\rm d^3r}
\end{equation}

with

\begin{eqnarray}
\rho_c(\vec r)=-\sum_{\mu \in c}\sum_{\vec g,\nu}P_{\nu\vec g\mu\vec 0}
\phi_{\mu}(\vec r-\vec A_{\mu})\phi_{\nu}(\vec r-\vec A_{\nu}-\vec g)
\end{eqnarray}

and

\begin{eqnarray}
\rho_c^{\rm model}(\vec r)
=\sum_{l=0}^{L} \sum_{m=-l}^{l} \eta_l^m(\rho_c;\vec A_c)
\delta_l^m(\vec A_c,\vec r)
\end{eqnarray}

and

\begin{eqnarray}
\delta_{l}^{m}(\vec {A_c},\vec r)=\lim_{\alpha \rightarrow \infty}
Z_{l}^{m}(\hat{\vec A_c})\Lambda(\alpha,\vec r-\vec A_c,0,0,0)
\end{eqnarray}

$c$ is chosen as the set of basis functions sited at center $\vec A_c$.

\subsection{McMurchie-Davidson algorithm}

\label{McMurchieAlgorithm}

In this section, we indicate how the integrals are evaluated.
This is done with the McMurchie-Davidson \cite{McMurchieDavidson} algorithm.
In this formalism, the product of two Gaussian type functions is expanded
at an intermediate center 
in terms of Hermite Gaussian type functions. 

\begin{eqnarray}
\label{S1S2lambda}
S(\alpha,\vec r-\vec A,n,l,m)S(\beta,\vec r-\vec B,\tilde n,\tilde l,\tilde m)=\nonumber \\
\sum_{t,u,v} E(n,l,m,\tilde n,\tilde l,\tilde m,t,u,v)
\Lambda(\gamma,\vec r-\vec P,t,u,v)
\end{eqnarray}

with $\gamma=\alpha+\beta$ and 
$\vec P=\frac{\alpha \vec A+\beta\vec B}{\alpha+\beta}$.
$E$ also depends on $\alpha, \ \beta$ 
and the distance $\vec B-\vec A$; however,
the dependence on these parameters is suppressed in the notation.

This makes a very efficient evaluation of integrals feasible. 
The starting point $E(0,0,0,0,0,0,0,0,0)=\exp(-\frac{\alpha\beta}{\alpha+\beta}
|\vec B-\vec A|^2)$ can be derived from the Gaussian product rule
\cite{Boys,McWeeny}:

\begin{equation}
\exp(-\alpha |\vec r-\vec A|^2)\exp(-\beta |\vec r-\vec B|^2)=
\exp\bigg(-\frac{\alpha\beta}{\alpha+\beta}|\vec B-\vec A|^2\bigg)
\exp\bigg(-(\alpha+\beta)
\bigg|\vec r -\frac {\alpha\vec A + \beta \vec B}{\alpha+\beta}
\bigg|^2\bigg)
\end{equation}

The 
coefficients $E$ can be generated by recursion relations 
\cite{McMurchieDavidson,VicNATO}. They are zero for the case 
$t+u+v>2n +2 \tilde n+l+\tilde l$ and for all negative values of $t,u$ or $v$.

\section{Calculation of derivatives}

\label{Calculationofderivatives}

\subsection{Gradients within the McMurchie-Davidson algorithm}

The evaluation of gradients of the integrals is closely related to
the evaluation of the integrals themselves. 
All the integrals can be expressed
with the help of the 
$E$-coefficients\cite{McMurchieDavidson,VicNATO,VicCoulomb,Dovesi1983}.
In the following we show how derivatives of the integrals can be expressed
in a similar way with Hermite Gaussian type functions. Starting from
equation \ref{S1S2lambda}, we obtain

\begin{eqnarray}
\label{EGequation} 
& & \frac{\partial}{\partial A_x} 
\Big(S(\alpha,\vec r-\vec A,n,l,m)S(\beta,\vec r-\vec B,\tilde n,\tilde l,\tilde m)\Big)
=\nonumber \\ & & 
\frac{\partial}{\partial A_x}\sum_{t,u,v} E(n,l,m,\tilde n,\tilde l,\tilde m,t,u,v)
\Lambda(\gamma,\vec r-\vec P,t,u,v)=\nonumber \\ & & 
\sum_{t,u,v}
\bigg(\Big
(\frac{\partial}{\partial A_x}E(n,l,m,\tilde n,\tilde l,\tilde m,t,u,v)\Big)
\Lambda(\gamma,\vec r-\vec P,t,u,v)+ \nonumber \\ & &
\frac{\alpha}{\alpha+\beta}E(n,l,m,\tilde n,\tilde l,\tilde m,t,u,v)
\Lambda(\gamma,\vec r-\vec P,t+1,u,v)\bigg)=\nonumber \\ & & 
\sum_{t,u,v}\Big(\frac{\partial}{\partial A_x}E(n,l,m,\tilde n,\tilde l,\tilde m,t,u,v)
+\frac{\alpha}{\alpha+\beta}E(n,l,m,\tilde n,\tilde l,\tilde m,t-1,u,v)
\Big)\Lambda(\gamma,\vec r-\vec P,t,u,v))=\nonumber \\ & & 
\sum_{t,u,v}G_x^A(n,l,m,\tilde n,\tilde l,\tilde m,t,u,v)\Lambda(\gamma,\vec r-\vec P,t,u,v)
\end{eqnarray}

Therefore, the gradients can be obtained in a quite similar way as the
integrals. Instead of the $E$-coefficients, the coefficients $G_x^A$
(and, for the other derivatives, 
$G_y^A$, $G_z^A$, $G_x^B$, $G_y^B$ and $G_z^B$) have to be used, as defined
in equation \ref{EGequation}.
We obtain the following relation from equation \ref{EGequation}:

\begin{eqnarray}
\label{EGstart}
G_x^A(n,l,m,\tilde n,\tilde l,\tilde m,t,u,v)=\nonumber \\
\frac{\partial}{\partial A_x}E(n,l,m,\tilde n,\tilde l,\tilde m,t,u,v)+\nonumber \\
\frac{\alpha}{\alpha+\beta}E(n,l,m,\tilde n,\tilde l,\tilde m,t-1,u,v)
\end{eqnarray}

We could thus derive the $G$-coefficients from the $E$-coefficients.
However, a more convenient way would be to have a recursion relation
similar to the $E$-coefficients.
Indeed, these relations can be obtained in an analogous way
\cite{Vicbook}. We give the relations here for the case of
complex spherical Gaussian type functions; a transformation to
real spherical Gaussian type functions is possible along the lines
given in ref. \onlinecite{CRYSTALbuch}.
The starting point can be obtained from equation \ref{EGstart} and
the definition of $E(0,0,0,0,0,0,0,0,0)$ given in section 
\ref{McMurchieAlgorithm}:

\begin{eqnarray}
G_x^A(0,0,0,0,0,0,0,0,0)=\nonumber \\
\frac{\partial}{\partial A_x}E(0,0,0,0,0,0,0,0,0)+\nonumber \\
\frac{\alpha}{\alpha+\beta}E(0,0,0,0,0,0,-1,0,0)=\nonumber \\
2\frac{\alpha\beta}{\alpha+\beta}(B_x-A_x)E(0,0,0,0,0,0,0,0,0) \nonumber\\
{\rm and}\hspace{5cm} \nonumber \\
G_x^A(0,0,0,0,0,0,1,0,0)=\nonumber \\
\frac{\partial}{\partial A_x}E(0,0,0,0,0,0,1,0,0)+\nonumber \\
\frac{\alpha}{\alpha+\beta}E(0,0,0,0,0,0,0,0,0)=\nonumber \\
\frac{\alpha}{\alpha+\beta}E(0,0,0,0,0,0,0,0,0)
\end{eqnarray}

All the other $G_x^A(0,0,0,0,0,0,t,u,v)$ are zero. Similarly, we obtain

\begin{eqnarray}
G_y^A(0,0,0,0,0,0,0,1,0)=\nonumber \\
G_z^A(0,0,0,0,0,0,0,0,1)=\nonumber \\
G_x^A(0,0,0,0,0,0,1,0,0)\nonumber \\
{\rm and}\hspace{5cm} \nonumber \\
G_y^A(0,0,0,0,0,0,0,0,0)=\nonumber \\
2\frac{\alpha\beta}{\alpha+\beta}(B_y-A_y)E(0,0,0,0,0,0,0,0,0)\nonumber \\
{\rm and}\hspace{5cm}\nonumber \\
G_z^A(0,0,0,0,0,0,0,0,0)=\nonumber \\
2\frac{\alpha\beta}{\alpha+\beta}(B_z-A_z)E(0,0,0,0,0,0,0,0,0)
\end{eqnarray}

Recursion relations for the $G$-coefficients can be derived using
similar arguments as for the $E$-coefficients\cite{Vicbook}. There exist 
 recursion relations to generate $E(n+1,l,m,\tilde n,\tilde l,\tilde m,t,u,v)$,
$E(n,l+1,m,\tilde n,\tilde l,\tilde m,t,u,v)$, $E(n,l+1,l+1,\tilde n,\tilde l,\tilde m,t,u,v)$
and $E(n,-l-1,-l-1,\tilde n,\tilde l,\tilde m,t,u,v)$. Recursions are now necessary
for $G_x^A$, $G_y^A$ and $G_z^A$.

\subsubsection{Recursion in $l$ and $m$}

With $S(\alpha,\vec r-\vec A,n,l+1,l+1)=
(2l+1)\big((x-A_x) +{\rm i}(y-A_y)\big)S(\alpha,\vec r-\vec A,n,l,l)$,
we obtain:

\begin{eqnarray} & &
S(\alpha,\vec r-\vec A,n,l+1,l+1)S(\beta,\vec r-\vec B,\tilde n,\tilde l,\tilde m)=\nonumber \\ & & 
(2l+1)\sum_{t,u,v} E(n,l,l,\tilde n,\tilde l,\tilde m,t,u,v)\big((x-A_x) +{\rm i}(y-A_y)\big)
\Lambda(\gamma,\vec r-\vec P,t,u,v)=\nonumber \\ & & 
\sum_{t,u,v} E(n,l+1,l+1,\tilde n,\tilde l,\tilde m,t,u,v)\Lambda(\gamma,\vec r-\vec P,t,u,v)
\end{eqnarray}

In the case of gradients, we obtain:

\begin{eqnarray}
\label{lmequation}
& & 
\frac{\partial}{\partial A_x}
S(\alpha,\vec r-\vec A,n,l+1,l+1)S(\beta,\vec r-\vec B,\tilde n,\tilde l,\tilde m)=\nonumber \\ & & 
(2l+1)\frac{\partial}{\partial A_x}
\bigg(\sum_{t,u,v} E(n,l,l,\tilde n,\tilde l,\tilde m,t,u,v)\big((x-A_x) +{\rm i}(y-A_y)\big)
\Lambda(\gamma,\vec r-\vec P,t,u,v)\bigg)=\nonumber \\ & &
(2l+1)\Bigg(
\sum_{t,u,v} -E(n,l,l,\tilde n,\tilde l,\tilde m,t,u,v)\Lambda(\gamma,\vec r-\vec P,t,u,v)
+\nonumber \\ & & 
\big((x-A_x) +{\rm i}\big(y-A_y))\frac{\partial}{\partial A_x}
\bigg(E(n,l,l,\tilde n,\tilde l,\tilde m,t,u,v)\Lambda(\gamma,\vec r-\vec P,t,u,v)\bigg)\Bigg)
=\nonumber \\ & &
(2l+1)\Bigg(
\sum_{t,u,v} -E(n,l,l,\tilde n,\tilde l,\tilde m,t,u,v)\Lambda(\gamma,\vec r-\vec P,t,u,v)
+\nonumber \\ & &
\big((x-A_x) +{\rm i}(y-A_y)\big)
G_x^A(n,l,l,\tilde n,\tilde l,\tilde m,t,u,v)\Lambda(\gamma,\vec r-\vec P,t,u,v)\Bigg)=
\nonumber\\ & &
\sum_{t,u,v}G_x^A(n,l+1,l+1,\tilde n,\tilde l,\tilde m,t,u,v)\Lambda(\gamma,\vec r-\vec P,t,u,v)
\end{eqnarray}

By substituting the recursion relation for Hermite polynomials

\begin{eqnarray}
(x-P_x)\Lambda(\gamma,\vec r-\vec P,t,u,v)=\frac{1}{2\gamma}
\Lambda(\gamma,\vec r-\vec P,t+1,u,v)+t \Lambda(\gamma,\vec r-\vec P,t-1,u,v)
\end{eqnarray}

in the penultimate expression of equation \ref{lmequation}, we obtain:

\begin{eqnarray}
(2l+1)\sum_{t,u,v} -E(n,l,l,\tilde n,\tilde l,\tilde m,t,u,v)\Lambda(\gamma,\vec r-\vec P,t,u,v)
+\nonumber \\
\bigg(\frac{1}{2\gamma}\Lambda(\gamma,\vec r-\vec P,t+1,u,v)+\nonumber \\
t\Lambda(\gamma,\vec r-\vec P,t-1,u,v)+\nonumber \\
(P_x-A_x)\Lambda(\gamma,\vec r-\vec P,t,u,v)+\nonumber \\
{\rm i}\Big(\frac{1}{2\gamma}\Lambda(\gamma,\vec r-\vec P,t,u+1,v)+\nonumber \\
u\Lambda(\gamma,\vec r-\vec P,t,u-1,v)+\nonumber \\
(P_y-A_y)\Lambda(\gamma,\vec r-\vec P,t,u,v)\Big)\bigg)
G_x^A(n,l,l,\tilde n,\tilde l,\tilde m,t,u,v)
\end{eqnarray}

From this, we deduce the following relation:

\begin{eqnarray}
G_x^A(n,l+1,l+1,\tilde n,\tilde l,\tilde m,t,u,v)=\nonumber \\
(2l+1)\bigg(-E(n,l,l,\tilde n,\tilde l,\tilde m,t,u,v)+\nonumber \\
\frac{1}{2\gamma}G_x^A(n,l,l,\tilde n,\tilde l,\tilde m,t-1,u,v)+\nonumber \\
(t+1)G_x^A(n,l,l,\tilde n,\tilde l,\tilde m,t+1,u,v)+\nonumber \\
(P_x-A_x)G_x^A(n,l,l,\tilde n,\tilde l,\tilde m,t,u,v)+\nonumber \\
{\rm i}\Big(\frac{1}{2\gamma}G_x^A(n,l,l,\tilde n,\tilde l,\tilde m,t,u-1,v)+\nonumber \\
(u+1)G_x^A(n,l,l,\tilde n,\tilde l,\tilde m,t,u+1,v)+\nonumber \\
(P_y-A_y)G_x^A(n,l,l,\tilde n,\tilde l,\tilde m,t,u,v)\Big)\bigg)
\end{eqnarray}

In an analogous way, we obtain the following
recursion relation:

\begin{eqnarray}
G_x^A(n,l+1,-l-1,\tilde n,\tilde l,\tilde m,t,u,v)=\nonumber \\
(2l+1)\bigg(-E(n,l,-l,\tilde n,\tilde l,\tilde m,t,u,v)+\nonumber \\
\frac{1}{2\gamma}G_x^A(n,l,-l,\tilde n,\tilde l,\tilde m,t-1,u,v)+\nonumber \\
(t+1)G_x^A(n,l,-l,\tilde n,\tilde l,\tilde m,t+1,u,v)+\nonumber \\
(P_x-A_x)G_x^A(n,l,-l,\tilde n,\tilde l,\tilde m,t,u,v)-\nonumber \\
{\rm i}\Big(\frac{1}{2\gamma}G_x^A(n,l,-l,\tilde n,\tilde l,\tilde m,t,u-1,v)+\nonumber \\
(u+1)G_x^A(n,l,-l,\tilde n,\tilde l,\tilde m,t,u+1,v)+\nonumber \\
(P_y-A_y)G_x^A(n,l,-l,\tilde n,\tilde l,\tilde m,t,u,v)\Big)\bigg)
\end{eqnarray}

\subsubsection{Recursion in n}

With
$S(\alpha,\vec r-\vec A,n+1,l,m)=
  |\vec r-\vec A|^2S(\alpha,\vec r-\vec A,n,l,m)$
the following relation can be derived \cite{Vicbook}:

\begin{eqnarray}
G_x^A(n+1,l,m,\tilde n,\tilde l,\tilde m,t,u,v)=\nonumber \\
\frac{1}{4\gamma^2}\Big(G_x^A(n,l,m,\tilde n,\tilde l,\tilde m,t-2,u,v)+\nonumber \\
G_x^A(n,l,m,\tilde n,\tilde l,\tilde m,t,u-2,v)+\nonumber \\
G_x^A(n,l,m,\tilde n,\tilde l,\tilde m,t,u,v-2)\Big)+\nonumber \\
\frac{1}{\gamma}\Big((P_x-A_x)G_x^A(n,l,m,\tilde n,\tilde l,\tilde m,t-1,u,v)-\nonumber \\
E(n,l,m,\tilde n,\tilde l,\tilde m,t-1,u,v)+\nonumber \\
(P_y-A_y)G_x^A(n,l,m,\tilde n,\tilde l,\tilde m,t,u-1,v)+\nonumber \\
(P_z-A_z)G_x^A(n,l,m,\tilde n,\tilde l,\tilde m,t,u,v-1)\Big)+\nonumber \\
\Big(|\vec P-\vec A|^2+\frac{(t+u+v+\frac{3}{2})}{\gamma}\Big)
G_x^A(n,l,m,\tilde n,\tilde l,\tilde m,t,u,v)-\nonumber \\
2(P_x-A_x)E(n,l,m,\tilde n,\tilde l,\tilde m,t,u,v)+\nonumber \\
2\Big((P_x-A_x)(t+1)G_x^A(n,l,m,\tilde n,\tilde l,\tilde m,t+1,u,v)-\nonumber \\
(t+1)E(n,l,m,\tilde n,\tilde l,\tilde m,t+1,u,v)+\nonumber \\
(P_y-A_y)(u+1)G_x^A(n,l,m,\tilde n,\tilde l,\tilde m,t,u+1,v)+\nonumber \\
(P_z-A_z)(v+1)G_x^A(n,l,m,\tilde n,\tilde l,\tilde m,t,u,v+1)\Big)+\nonumber \\
(t+2)(t+1)G_x^A(n,l,m,\tilde n,\tilde l,\tilde m,t+2,u,v)+\nonumber \\
(u+2)(u+1)G_x^A(n,l,m,\tilde n,\tilde l,\tilde m,t,u+2,v)+\nonumber \\
(v+2)(v+1)G_x^A(n,l,m,\tilde n,\tilde l,\tilde m,t,u,v+2)
\end{eqnarray}

\subsubsection{Recursion in l}

Using that
$S(\alpha,\vec r-\vec A,n,l+1,m)=
 ((2l+1)(z-A_z)S(\alpha,\vec r-\vec A,n,l,m)-
 |\vec r-\vec A|^2(l+|m|)S(\alpha,\vec r-\vec A,n,l-1,m))/(l-|m|+1)$
the following relation can be derived \cite{Vicbook}:

\begin{eqnarray}
G_x^A(n,l+1,m,\tilde n,\tilde l,\tilde m,t,u,v)=\nonumber \\
\Bigg((2l+1)\bigg(\frac{G_x^A(n,l,m,\tilde n,\tilde l,\tilde m,t,u,v-1)}{2\gamma}+\nonumber \\
(P_z-A_z)G_x^A(n,l,m,\tilde n,\tilde l,\tilde m,t,u,v)+\nonumber \\
(v+1)G_x^A(n,l,m,\tilde n,\tilde l,\tilde m,t,u,v+1)\bigg)-\nonumber \\
(l+|m|)\bigg(\Big(G_x^A(n,l-1,m,\tilde n,\tilde l,\tilde m,t-2,u,v)+\nonumber \\
G_x^A(n,l-1,m,\tilde n,\tilde l,\tilde m,t,u-2,v)+\nonumber \\
G_x^A(n,l-1,m,\tilde n,\tilde l,\tilde m,t,u,v-2)\Big)/(2\gamma)^2+\nonumber \\
\Big((P_x-A_x)G_x^A(n,l-1,m,\tilde n,\tilde l,\tilde m,t-1,u,v)-\nonumber \\
E(n,l-1,m,\tilde n,\tilde l,\tilde m,t-1,u,v)+\nonumber \\
(P_y-A_y)G_x^A(n,l-1,m,\tilde n,\tilde l,\tilde m,t,u-1,v)+\nonumber \\
(P_z-A_z)G_x^A(n,l-1,m,\tilde n,\tilde l,\tilde m,t,u,v-1)\Big)/\gamma+\nonumber \\ 
\Big(|\vec P-\vec A|^2+\frac{(t+u+v+\frac{3}{2})}{\gamma}\Big)
G_x^A(n,l-1,m,\tilde n,\tilde l,\tilde m,t,u,v)-\nonumber \\
2(P_x-A_x)E(n,l-1,m,\tilde n,\tilde l,\tilde m,t,u,v)+\nonumber \\
2\Big((P_x-A_x)(t+1)G_x^A(n,l-1,m,\tilde n,\tilde l,\tilde m,t+1,u,v)-\nonumber \\
(t+1)E(n,l-1,m,\tilde n,\tilde l,\tilde m,t+1,u,v)+\nonumber \\
(P_y-A_y)(u+1)G_x^A(n,l-1,m,\tilde n,\tilde l,\tilde m,t,u+1,v)+\nonumber \\
(P_z-A_z)(v+1)G_x^A(n,l-1,m,\tilde n,\tilde l,\tilde m,t,u,v+1)\Big)+\nonumber \\
(t+2)(t+1)G_x^A(n,l-1,m,\tilde n,\tilde l,\tilde m,t+2,u,v)+\nonumber \\
(u+2)(u+1)G_x^A(n,l-1,m,\tilde n,\tilde l,\tilde m,t,u+2,v)+\nonumber \\
(v+2)(v+1)G_x^A(n,l-1,m,\tilde n,\tilde l,\tilde m,t,u,v+2)\bigg)\Bigg)/(l-|m|+1)
\end{eqnarray}

The recursion relations for the $G_x^A$-coefficients 
are similar to those for the $E$-coefficients
\cite{VicNATO}, with the extra terms arising from the derivative
$\frac{\partial}{\partial A_x}(x-A_x)$ for the recursion in $l$ and $m$,
and from the derivative $\frac{\partial}{\partial A_x}|\vec r-\vec A|^2$ for
the recursion in $l$ and the recursion in $n$.

This can similarly be done for $G_y^A$ and $G_z^A$.
Finally, all $G$-coefficients are zero for the case 
$t+u+v>2n+2\tilde n+l+\tilde l+1$ and for all negative values of $t,u$ or $v$.

\subsection{Gradients with respect to other centers}

To obtain the derivatives with respect to center $B$, 
the following relation is used \cite{HelgakerTaylor} 
(note $\vec R=\vec A-\vec B$):

\begin{eqnarray}
\frac{\partial}{\partial A_x}=\frac{\partial}{\partial P_x}\frac{\partial P_x}
{\partial A_x}+\frac{\partial}{\partial R_x}
\frac{\partial R_x}{\partial A_x}=\nonumber \\
\frac{\alpha}{\gamma}\frac{\partial}{\partial P_x}+
\frac{\partial}{\partial R_x}\nonumber \\
\end{eqnarray}

and

\begin{eqnarray}
\frac{\partial}{\partial B_x}=\frac{\partial}{\partial P_x}
\frac{\partial P_x}{\partial B_x}+\frac{\partial}{\partial R_x}
\frac{\partial R_x}{\partial B_x}=\nonumber \\
\frac{\beta}{\gamma}\frac{\partial}{\partial P_x}-
\frac{\partial}{\partial R_x}\nonumber \\
\end{eqnarray}

Therefore,

\begin{eqnarray}
\frac{\partial}{\partial A_x}+\frac{\partial}{\partial B_x}=
\frac{\partial}{\partial P_x}
\end{eqnarray}

This means that 

\begin{eqnarray}
\frac{\partial}{\partial B_x}
\sum_{t,u,v}E(n,l,m,\tilde n,\tilde l,\tilde m,t,u,v)\Lambda(\gamma,\vec r-\vec P,t,u,v)
=\nonumber \\
\left(\frac{\partial}{\partial P_x}-\frac{\partial}{\partial A_x}\right)
\sum_{t,u,v}E(n,l,m,\tilde n,\tilde l,\tilde m,t,u,v)\Lambda(\gamma,\vec r-\vec P,t,u,v)
\end{eqnarray}

Applying equation \ref{EGequation}, we obtain:

\begin{eqnarray} & & 
\sum_{t,u,v}G_x^B(n,l,m,\tilde n,\tilde l,\tilde m,t,u,v)\Lambda(\gamma,\vec r-\vec P,t,u,v)=\nonumber \\  &  &
\sum_{t,u,v}E(n,l,m,\tilde n,\tilde l,\tilde m,t,u,v)\Lambda(\gamma,\vec r-\vec P,t+1,u,v) \nonumber \\ & &
-\sum_{t,u,v}G_x^A(n,l,m,\tilde n,\tilde l,\tilde m,t,u,v)\Lambda(\gamma,\vec r-\vec P,t,u,v)
\end{eqnarray}

Finally, we conclude:

\begin{eqnarray}
G_x^B(n,l,m,\tilde n,\tilde l,\tilde m,t,u,v)=E(n,l,m,\tilde n,\tilde l,\tilde m,t-1,u,v)
-G_x^A(n,l,m,\tilde n,\tilde l,\tilde m,t,u,v)
\end{eqnarray}

and similar for derivatives with respect to $y$ and $z$ direction.
All the integrals can be expressed with the help of the $E$-coefficients.
Taking the derivative therefore reduces to replacing $E$ with the
corresponding $G_x^A$, $G_y^A$, etc coefficients.

As an example, we show how overlap integral and gradient are obtained.
First, the overlap is computed as follows:

\begin{eqnarray}
&  & \int S(\alpha,\vec r - \vec A,n,l,m)
     S(\beta,\vec r-\vec B,\tilde n,\tilde l,\tilde m){\rm d^3r}= \nonumber\\
& & \int\sum_{t,u,v} E(n,l,m,\tilde n,\tilde l,\tilde m,t,u,v) 
\Lambda(\gamma,\vec r-\vec P,t,u,v)
{\rm d^3r}= \nonumber\\ & & 
E(n,l,m,\tilde n,\tilde l,\tilde m,0,0,0)\left(\frac{\pi}{\gamma}\right)^{\frac {3}{2}}
\end{eqnarray}

We have used that 
\begin{eqnarray}
\int \Lambda(\gamma,\vec r,t,u,v){\rm d^3r}=\left(\frac{\pi}{\gamma}\right)
^{\frac{3}{2}}\delta_{t0}\delta_{u0}\delta_{v0}
\end{eqnarray}
because of the orthogonality of the Hermite Gaussian type functions
($\delta_{t0}$ is the Kronecker delta).

The gradient is computed similarly:

\begin{eqnarray} 
& & \frac{\partial}{\partial A_x}
\int S(\alpha,\vec r - \vec A,n,l,m)
     S(\beta,\vec r-\vec B,\tilde n,\tilde l,\tilde m){\rm d^3r}= \nonumber\\
& & \int\sum_{t,u,v} G_x^A(n,l,m,\tilde n,\tilde l,\tilde m,t,u,v) 
\Lambda(\gamma,\vec r-\vec P,t,u,v)
{\rm d^3r}=  \nonumber\\ & & 
G_x^A(n,l,m,\tilde n,\tilde l,\tilde m,0,0,0)\left(\frac{\pi}{\gamma}\right)^{\frac {3}{2}}
\end{eqnarray}

In our implementation, we therefore compute the gradient of the two Gaussians
 which
are associated with the integrals, by replacing $E$-coefficients
 with $G$-coefficients.
As a consequence, if an operator, which might appear in the integral,
has a nonvanishing derivative
(such as, for example, the nuclear attraction), this must be taken into
account additionally. This derivative with respect to the third center
can be obtained by applying translational invariance with respect to
a simultaneous uniform translation  of the three centers.
In the case of bielectronic integrals, products with two $E$-coefficients
appear. Obviously, when differentiating, the corresponding rules of
differentiating a product must be applied and two derivative terms
appear, each of them consisting of a product of one set of $E-$ and one set
of $G-$coefficients. 
Finally, the nuclear-nuclear term must be differentiated which is trivial.

It is interesting to compare this implementation with that of the Namur
group \cite{Jacquemin} where also gradients within the McMurchie-Davidson
algorithm are computed. Whereas our scheme computes the derivatives of the
two Gaussians appearing in the integral 
and a possibly necessary derivative of an operator is obtained
by applying translational invariance, the alternative 
implementation\cite{Jacquemin}
relies on explicitly computing derivatives of $E$-coefficients and
of the auxiliary function \cite{McMurchieDavidson} $R_{t,u,v}$ appearing
in the integrals.

\section{Total energy}

\label{FMexplizit}

The total energy consists of kinetic energy, Coulomb energy
(nuclear-nuclear repulsion, nuclear-electron attraction and electron-electron
repulsion), and exchange energy. We assume that
all the orbitals are either empty or doubly occupied.

\subsection{Kinetic energy}
The kinetic energy of the electrons is obtained as:

\begin{eqnarray}
E^{\rm kinetic}=\sum_{\vec g,\mu,\nu}P_{\nu\vec g\mu\vec 0}
\int {\phi_{\mu}(\vec r-\vec A_{\mu})\bigg( -\frac{1}{2}\Delta_{\vec r}\bigg)
\phi_{\nu}(\vec r-\vec A_{\nu}-\vec g){\rm d^3r}}=\nonumber \\
\sum_{\vec g,\mu,\nu}P_{\nu\vec g\mu\vec 0}
T_{\mu\vec 0\nu\vec g}
\end{eqnarray}

\subsection{Exchange energy}

The exchange energy is obtained as:

\begin{eqnarray}
& &
E^{\rm exch-el}=\nonumber \\ & &
-\frac{1}{4} \sum_{\vec g,\mu,\nu}P_{\nu\vec g\mu\vec 0}
\sum_{\vec h,\vec n,\tau,\sigma}P_{\sigma\vec h+\vec n \tau\vec n}
\nonumber \\ & &
\int \frac {\phi_{\mu}(\vec r-\vec A_{\mu})
\phi_{\tau}(\vec r-\vec A_{\tau}-\vec n)
\phi_{\nu}(\vec r \: '-\vec A_{\nu}-\vec g)\phi_{\sigma}
(\vec r \: ' - \vec A_{\sigma}
-\vec h-\vec n)}
{|\vec r-\vec r \: '|}{\rm d^3r\: d^3r'}=\nonumber \\ & &
-\frac{1}{4}\sum_{\vec g,\mu,\nu}P_{\nu\vec g\mu\vec 0}
\sum_{\vec h,\tau,\sigma}P_{\sigma\vec h\tau\vec 0}
X_{\mu\vec 0\nu\vec g\tau\vec 0\sigma\vec h}
\end{eqnarray}

where we have exploited translational invariance of the density matrix
with respect to direct lattice vectors $\vec n$:
$P_{\sigma\vec h+\vec n \tau\vec n}=P_{\sigma\vec h\tau\vec 0}$.

We can define a Fock operator for the exchange energy which is

\begin{eqnarray}
F_{\mu\vec 0\nu\vec g}^{\rm exch-el}=-\frac{1}{2}
\sum_{\vec h,\tau,\sigma}P_{\sigma\vec h\tau\vec 0}
X_{\mu\vec 0\nu\vec g\tau\vec 0\sigma\vec h}
\end{eqnarray}

\subsection{Coulomb energy}

Both kinetic energy and exchange energy must converge independently.
However, a separation of the contributions to the Coulomb energy is not
possible: for example, in a one dimensional periodic system with lattice
constant $a$, and $n$ being an index enumerating the cells, the 
electron-electron interaction per unit cell would have contributions like:

\begin{eqnarray}
\sum_{n=1}^{\infty} \frac{1}{n a}
\end{eqnarray}

This sum is divergent (similarly in two and three dimensions).
Therefore, in CRYSTAL a scheme based on the Ewald method is used to
sum the interactions \cite{Harris,VicCoulomb}.
We only quote the results for the individual contributions:

\subsubsection{Nuclear-nuclear repulsion:}

\begin{eqnarray}
E^{\rm NN}=\frac{1}{2}\sum_{a,b}Z_a Z_b A(\vec A_b-\vec A_a)
\end{eqnarray}

\subsubsection{Nuclear-electron attraction:}

The energy $\frac{1}{2}E_{\rm NE}$, which is the Ewald energy
of the nuclei in the primitive unit cell with the all 
the electrons of all cells, is the same as the energy $\frac{1}{2}E_{\rm EN}$,
which is the Ewald energy of the electrons of the primitive unit cell with
all the nuclei in all cells, as long as no
approximations are introduced\cite{Dovesi1983}. CRYSTAL uses the following
expression as the sum of these interactions:

\begin{eqnarray}
E^{\rm coul-nuc}=\sum_{\vec g,\mu,\nu} 
P_{\nu\vec g\mu\vec 0}F^{\rm coul-nuc}_{\mu\vec 0\nu\vec g}
\end{eqnarray}

with the Fock matrix $F^{\rm coul-nuc}_{\mu\vec 0\nu\vec g}$ containing the
nuclear-electron contributions defined as

\begin{eqnarray}
F^{\rm coul-nuc}_{\mu\vec 0\nu\vec g}=
-\sum_{a}Z_a\int\phi_{\mu}(\vec r-\vec A_{\mu})
\phi_{\nu}(\vec r-A_{\nu}-\vec g)A(\vec r-\vec A_{a}){\rm d^3r}
\end{eqnarray}

\subsubsection{Electron-electron repulsion:}

\begin{eqnarray}
E^{\rm coul-el}
=\frac{1}{2}\sum_{\vec g,\mu,\nu} P_{\nu\vec g\mu\vec 0}
F^{\rm coul-el}_{\mu\vec 0\nu\vec g}
\end{eqnarray}

with the Fock matrix $F^{\rm coul-el}_{\mu\vec 0\nu\vec g}$ containing the
electron-electron contributions defined as

\begin{eqnarray}
F^{\rm coul-el}_{\mu\vec 0\nu\vec g}=
-QS_{\mu\vec 0\nu\vec g}+\sum_{\vec h,\tau,\sigma}P_{\sigma\vec h\tau\vec 0}
 C_{\mu\vec 0\nu\vec g \tau\vec 0 \sigma\vec h}
-\sum_c \sum_{l=0}^{L}\sum_{m=-l}^{l}\eta_l^m(\rho_c;\vec A_c)
M_{l\mu\vec 0\nu\vec gc}^m
\end{eqnarray}

\subsection{Total energy}

Finally, the total energy can be expressed as

\begin{eqnarray} & & 
E^{\rm total}=E^{\rm kinetic}+E^{\rm NN}+E^{\rm coul-nuc}+E^{\rm coul-el}
+E^{\rm exch-el}=\nonumber \\ & & 
=E^{\rm NN}+\sum_{\vec g,\mu,\nu}P_{\nu\vec g\mu\vec 0}T_{\mu\vec 0\nu\vec g}
\nonumber \\ & & 
-\sum_{\vec g,\mu,\nu} 
P_{\nu\vec g\mu\vec 0}\sum_{a}Z_a\int\phi_{\mu}(\vec r-\vec A_{\mu})
\phi_{\nu}(\vec r-\vec A_{\nu}-\vec g)A(\vec r-\vec A_{a}){\rm d^3r}
\nonumber \\ & & 
+\frac{1}{2}\sum_{\vec g,\mu,\nu} P_{\nu\vec g\mu\vec 0}
\bigg(-QS_{\mu\vec 0\nu\vec g}+\sum_{\vec h,\tau,\sigma}
P_{\sigma\vec h\tau\vec 0}
C_{\mu\vec 0\nu\vec g \tau\vec 0\sigma\vec h}
-\sum_c \sum_{l=0}^{L}\sum_{m=-l}^{l}\eta_l^m(\rho_c;\vec A_c)
M_{l\mu\vec 0\nu\vec gc}^m\bigg)\nonumber \\ & & 
-\frac{1}{4}\sum_{\vec g,\mu,\nu}P_{\nu\vec g\mu\vec 0}
\sum_{\vec h,\tau,\sigma}P_{\sigma\vec h\tau\vec 0}
X_{\mu\vec 0\nu\vec g\tau\vec 0\sigma\vec h}
\end{eqnarray}

The Fock operator used has the structure:

\begin{eqnarray}
F^{\rm total}_{\mu\vec 0\nu\vec g}=T_{\mu\vec 0\nu\vec g}+
F^{\rm coul-nuc}_{\mu\vec 0\nu\vec g}+F^{\rm coul-el}_{\mu\vec 0\nu\vec g}
+F^{\rm exch-el}_{\mu\vec 0\nu\vec g}
\end{eqnarray}

We note that this expression for the Fock operator would be exact
if we could guarantee that the penetration depth and screening was symmetric.
This would require that
$C_{\mu\vec 0\nu\vec g \tau\vec 0\sigma\vec h}=
C_{\tau\vec 0\sigma\vec h\mu\vec 0\nu\vec g}$ should always hold.
This, however, as aforementioned,
cannot be guaranteed because the truncation is
not necessarily 
symmetric.
In addition, the screening of the exchange interaction is not necessarily
symmetric. 
Therefore, an inaccuracy in the Fock operator will show up which will
be stronger the more asymmetric the truncation in the energy expression is. 

The total energy can be expressed as

\begin{equation}
E^{\rm total}=E^{\rm NN}+\sum_{\vec g,\mu,\nu}P_{\nu\vec g\mu\vec 0}
\bigg(T_{\mu\vec 0\nu\vec g}+F^{\rm coul-nuc}_{\mu\vec 0\nu\vec g}
+\frac{1}{2}(F^{\rm coul-el}_{\mu\vec 0\nu\vec g}
+F^{\rm exch-el}_{\mu\vec 0\nu\vec g})\bigg)
\end{equation}

and the Hartree-Fock equations become as in equation \ref{HFequation}.
In ref. \onlinecite{VicCoulomb}, it was pointed out 
that the quantity $QS_{\mu\vec 0\nu\vec g}$ 
can be removed from the Fock operator  which has been done in CRYSTAL.
This leads to eigenvalues shifted by $Q$ as we now use the modified equation

\begin{eqnarray}
\sum_{\nu}(F^{\rm total}_{\mu \nu}
(\vec k)+QS_{\mu \nu}(\vec k))a_{\nu i}(\vec k)=
\sum_{\nu}
S_{\mu \nu}(\vec k)a_{\nu i}(\vec k)(\epsilon_i(\vec k)+Q)
\end{eqnarray}

\section{Gradient of the total energy}

\label{Gradienttotenysection}

The force on the nuclei can be calculated similarly to the molecular case
\cite{Bratoz,Pulay}. The derivatives of all the integrals are necessary,
and  the derivative of the density matrix is expressed with the help
of the energy-weighted density matrix. One important assumption is that 

\begin{eqnarray}
\label{Bequivalence}
B_{\mu\vec 0\nu\vec g  \tau\vec n \sigma \vec n+\vec h}=
B_{\tau\vec n\sigma\vec n+\vec h  \mu\vec 0\nu\vec g}=
B_{\sigma\vec n+\vec h\tau\vec n  \nu\vec g\mu\vec 0}
\end{eqnarray}

 holds. Taking the derivative leads, for example, to terms like the following

\begin{eqnarray}
& & \frac{\partial}{\partial \vec A_i} \bigg(\sum_{\vec g,\mu,\nu}
P_{\nu\vec g\mu\vec 0}
\sum_{\vec h,\tau,\sigma}P_{\sigma\vec h\tau\vec 0}
C_{\mu\vec 0\nu\vec g \tau\vec 0\sigma\vec h}\bigg)= \nonumber \\ & &
\sum_{\vec g,\mu,\nu}\bigg(\frac{\partial}{\partial \vec A_i}
P_{\nu\vec g\mu\vec 0}\bigg)\sum_{\vec h,\tau,\sigma}P_{\sigma\vec h\tau\vec 0}
C_{\mu\vec 0\nu\vec g \tau\vec 0\sigma\vec h}+
\sum_{\vec g,\mu,\nu}P_{\nu\vec g\mu\vec 0}
\sum_{\vec h,\tau,\sigma}\bigg(\frac{\partial}{\partial \vec A_i}
P_{\sigma\vec h\tau\vec 0}\bigg)
C_{\mu\vec 0\nu\vec g \tau\vec 0\sigma\vec h}\nonumber \\ & &
+\sum_{\vec g,\mu,\nu}P_{\nu\vec g\mu\vec 0}
\sum_{\vec h,\tau,\sigma}P_{\sigma\vec h\tau\vec 0}
\bigg(\frac{\partial}{\partial \vec A_i}
C_{\mu\vec 0\nu\vec g \tau\vec 0\sigma\vec h}\bigg)
\end{eqnarray}

When equation \ref{Bequivalence}  holds, we rename the indices in the second
addend and obtain:

\begin{eqnarray}
2 \sum_{\vec g,\mu,\nu}\bigg(\frac{\partial}{\partial \vec A_i}
P_{\nu\vec g\mu\vec 0}\bigg)\sum_{\vec h,\tau,\sigma}
P_{\sigma\vec h\tau\vec 0}
C_{\mu\vec 0\nu\vec g \tau\vec 0\sigma\vec h}+
\sum_{\vec g,\mu,\nu}P_{\nu\vec g\mu\vec 0}
\sum_{\vec h,\tau,\sigma}P_{\sigma\vec h\tau\vec 0}
\bigg(\frac{\partial}{\partial \vec A_i}
C_{\mu\vec 0\nu\vec g \tau\vec 0\sigma\vec h}\bigg)
\end{eqnarray}

We derived the equation for the force this way
 although equation \ref{Bequivalence} does not always hold. Therefore, 
inaccuracies will appear when equation \ref{Bequivalence}  is  strongly
violated.
The full force is obtained as:

\begin{eqnarray}
\label{Forcegleichung} & &
\vec F_{A_i}=-\frac{\partial E^{\rm total}}{\partial \vec A_i}=\nonumber \\ & &
-\sum_{\vec g,\mu,\nu}P_{\nu\vec g\mu\vec 0}\frac{\partial 
T_{\mu\vec 0\nu\vec g}}
{\partial \vec A_i}
-\frac{\partial E^{\rm NN}}{\partial \vec A_i}\nonumber \\ & &
+\sum_{\vec g,\mu,\nu} P_{\nu\vec g\mu\vec 0}\sum_{a}Z_a
\frac{\partial}{\partial \vec A_i} \bigg[ \int\phi_{\mu}(\vec r-\vec A_{\mu})
\phi_{\nu}(\vec r-\vec A_{\nu}-\vec g)A(\vec r-\vec A_{a}){\rm d^3r}\bigg]
\nonumber \\ & &
-\frac{1}{2}\sum_{\vec g,\mu,\nu} P_{\nu\vec g\mu\vec 0}
\bigg\{-S_{\mu\vec 0\nu\vec g}
\frac{2\pi}{3V}
\sum_{c}
\sum_{\vec h,\sigma,\tau \in c}P_{\sigma \vec h \tau \vec 0}\nonumber \\ & &
\frac{\partial}{\partial \vec A_i} \int\bigg[
-\phi_{\tau}(\vec r-\vec A_{\tau})\phi_{\sigma}(\vec r-\vec A_{\sigma}-\vec h)
\nonumber \\ & &
+\sum_{l=0}^L \sum_{m=-l}^l\int \phi_{\tau}(\vec r \: '-\vec A{_\tau})
\phi_{\sigma}
(\vec r \: '-\vec A_{\sigma}-\vec h)
X_l^m(\vec r \: '-\vec A_c){\rm d^3r'}
\delta_l^m(\vec A_c,\vec r)\bigg]r^2 {\rm d^3r} \nonumber\\ & &
+\sum_{\tau,\sigma}P_{\sigma\vec h\tau\vec 0}
\frac{\partial C_{\mu\vec 0\nu\vec g \tau\vec 0\sigma\vec h}}
{\partial \vec A_i}\nonumber \\ & &
-\sum_c \sum_{l=0}^{L}\sum_{m=-l}^{l}\sum_{\vec h,\tau \in  c, \sigma}
P_{\sigma\vec h \tau \vec 0}
\frac{\partial}{\partial \vec A_i}\bigg[\int
\phi_{\tau}(\vec r-\vec A_{\tau})\phi_{\sigma}
(\vec r - \vec A_{\sigma}-\vec h)
X_l^m(\vec r-\vec {A_c}){\rm d^3r} \
 M_{l\mu\vec 0\nu\vec g c}^m\bigg]\bigg\}
\nonumber \\ & &
+\frac{1}{4}\sum_{\vec g,\mu,\nu} P_{\nu\vec g\mu\vec 0}
\sum_{\vec h,\tau,\sigma}P_{\sigma\vec h\tau\vec 0}
\frac{\partial X_{\mu\vec 0\nu\vec g\tau\vec 0\sigma\vec h}}
{\partial \vec A_i} \nonumber \\ & &
-\sum_{\vec g,\mu,\nu}
\frac{\partial S_{\mu\vec 0\nu\vec g}}{\partial \vec A_i}  \int_{BZ} 
 \exp({\rm i}\vec k\vec g)\sum_j 2a_{\nu j}(\vec k)a_{\mu j}^*(\vec k)
(\epsilon_j(\vec k)+Q)
\Theta(\epsilon_F-\epsilon_j(\vec k)-Q){\rm d^3k}
\end{eqnarray}

The last addend is the energy weighted density matrix; the integral
is over the first Brillouin zone.
It is worthwhile mentioning that the factor $P_{\nu\vec g\mu\vec 0}
S_{\mu\vec 0\nu\vec g}$ is equal to the number of electrons in the unit
cell and therefore its derivative with respect to $\vec A_i$ vanishes.
We note three important points:

\begin{itemize}
\item{Equation \ref{Forcegleichung} is correct for the exact solution
of the Hartree-Fock equations. Thus, in practice, a well converged solution
is necessary to achieve accurate forces.}
\item{The energy-weighted density matrix is $\vec k$-dependent. Therefore,
the accuracy of the forces will become dependent on the number of 
$\vec k$-points.}
\item{The derivation of equation \ref{Forcegleichung} assumes
that equation \ref{Bequivalence} holds. 
The treatment of the Coulomb series with finite penetration depth 
leads to an asymmetry associated with $ITOL2$ in the CRYSTAL
code. In addition, in the treatment
of the exchange series an asymmetry can be introduced if the 
screening parameters ($ITOL4$ and $ITOL5$ in the CRYSTAL
code) are chosen differently. Therefore, the choice of $ITOL2$, $ITOL4$ and 
$ITOL5$ will influence the accuracy of the gradients.}
\end{itemize}

\section{Results from test calculations}
\label{applications}

With a few examples, we want to illustrate the accuracy of the 
analytically computed gradients.

In table \ref{COmoleculeexact}, all the integrals are evaluated without
approximation and the analytical derivative agrees to five digits with
the numerical derivative. As the numerical derivative is only accurate
up to five digits, this is certainly satisfying.

In table \ref{COmoleculeitol} the variation in accuracy when penetration
depth and overlap criteria  are altered, is displayed.
As described in the article, lowering $ITOL2$
to low values leads to inaccuracies in the gradients.
Lowering only one of the parameters $ITOL4$ or $ITOL5$ also leads to 
inaccuracies, whereas lowering both to the same value gives an analytical
gradient which is consistent with the numerical gradient --- however,
as a value of 1 for $ITOL4$ and $ITOL5$ was chosen, energy and force are
completely different from calculations with reasonable values for $ITOL4$
and $ITOL5$. The parameter $ITOL1$, which selects the one-electron and 
Coulomb integrals according to
the overlap, can lead to numerical instabilities in the energy calculation
and inaccurate gradients 
when chosen too low, and should therefore be reasonably high. 
The parameter $ITOL3$ does not influence the accuracy of the gradients:
although when chosen much too low with a value of e.g. 1, numerical and
analytical gradient still agree.
The default values for the $ITOL$-parameters for the energy calculation
are 6, 6, 6, 6 and 12. In all
calculations performed so far, these default values did
not lead to serious errors for the gradients. 

Another example (table \ref{COdimensions}) is the CO molecule arranged as a
single molecule ("molecule"), as a molecule periodically repeated in
one dimension ("polymer"), in two dimensions ("slab") and in three
dimensions ("bulk"). The forces agree well, and it is demonstrated that
using stricter real-space truncation parameters improves the agreement.
The forces seem to be relatively insensitive to the number of sampling
points and
changing their number  changed the
error in the forces only slightly.

Finally, in table \ref{MgO421} we compare analytical and numerical derivatives
for MgO when moving the oxygen atom in x-direction which would correspond
to a longitudinal phonon. Again, agreement is to the order of
$10^{-5}\frac{E_h}{a_B}$ (default $ITOL$ parameters were used).

\section{Conclusion}
We presented the theory of analytic Hartree-Fock gradients for 
periodic systems. This has been implemented in the code CRYSTAL
which is to the best of our knowledge
 the first implementation of Hartree-Fock gradients in
systems periodic in 2 and 3 dimensions. The results are in 
excellent agreement with numerical derivatives.

Future directions will be the improvement of the efficiency of the code
(implementation of symmetry and various technical
improvements), derivatives with respect to the lattice vector, 
 as well as an extension to metallic systems.

\section{Acknowledgments}
The authors would like to acknowledge support from EPSRC grant
GR/K90661.

\newpage
\onecolumn

\begin{table}
\begin{center}
\caption{\label{COmoleculeexact}CO molecule, all tolerances high enough
that all the integrals are done without approximation. The carbon atom
is placed at (0 \AA, 0 \AA, 0 \AA), the oxygen atom at 
(0.8 \AA, 0.5 \AA, 0.4 \AA). To calculate the numerical force in 
x-direction, the
x-coordinate of the oxygen atom is changed (column 1). The Hartree-Fock
energy is displayed in column 2. The numerical force is determined
from 2 points at (0.8 $\pm \delta$) \AA \ and displayed in column 3. The
analytical derivative for x=0.8 \AA \ is displayed in column 4. Basis sets
of the size $[2s2p1d]$ were used.}
%\vspace{5mm}
\begin{tabular}{cccc} 
x-coordinate of oxygen & energy & numerical derivative 
& analytical 
derivative\\
& & (x-component) & (x-component) \\
\AA   & $E_h$               &  $E_h/a_B$ & $E_h/a_B$\\
0.799   & -1.107642201574E+02 & 0.376915 \\
0.7999  & -1.107648645627E+02 & 0.376913\\
0.79999 & -1.107649287000E+02 & 0.376914            \\
0.80000 & -1.107649358230E+02 &          & 0.37691274\\
0.80001 & -1.107649429453E+02 &\\
0.8001  & -1.107650070153E+02 &\\
0.801   & -1.107656446886E+02 &\\
\end{tabular}
\end{center}
\end{table}

\newpage
\begin{table}
\begin{center}
\caption{\label{COmoleculeitol} CO molecule, when varying the $ITOL$ 
parameters.
The first and
second parameter should be reasonably high, the fourth and fifth should
both be high or identical to obtain accurate forces.}
\begin{tabular}{ccc} 
$ITOL$ values  & numerical derivative  & analytical derivative \\
 & (x-component) & (x-component)\\
     &   $E_h/a_B$ & $E_h/a_B$\\
   6 6 6 6 12 & 0.37691 & 0.376913 \\
20 20 20 20 20 & 0.37691 & 0.376913 \\
4 20 20 20 20 & 0.37691 & 0.376912 \\
3 20 20 20 20 & 0.37747 & 0.377389 \\
2 20 20 20 20 & 0.38883 & 0.388260 \\
20 2 20 20 20 & 0.37684 & 0.376821 \\
20 1 20 20 20 & -7.40676 & -1.246409 \\
20 20 1 20 20 & -8.09240 & -8.09240 \\
20 20 20 1 20 & -0.51905 & -3.504654 \\
20 20 20 20 1 &  0.40892 & 0.359847\\
20 20 20 1  1 & -33.34823 & -33.348229 \\
\end{tabular}
\end{center}
\end{table}

\newpage
\begin{table}
\begin{center}
\caption{\label{COdimensions} CO molecule, arranged periodic in 
0 dimensions ("molecule"), in 1 dimension ("polymer"), in 2 dimensions ("slab")
and in 3 dimensions ("bulk"). We display the accuracy of the gradient
as a function of the number of sampling points, and as a function of
the $ITOL$ parameters.}

\begin{tabular}{ccccc} 
dimension & number of sampling points & $ITOL$ parameters 
& numerical derivative & analytical derivative \\
& & & (x-component) & (x-component)\\
  &   &            &   $E_h/a_B$ & $E_h/a_B$\\
0 & - & 6 6 6 6 12 & 0.37691 & 0.376913 \\ \\
1 & 5  & 6 6 6 6 12 & 0.37659 & 0.376633\\
1 & 5  & 6 9 6 12 12 & 0.37662 & 0.376647 \\ \\
2 & 34  & 6 6 6 6 12 & 0.37630 & 0.376335 \\ \\
3 & 8    & 6 6 6 6 12 & 0.37571 & 0.375742 \\
3 & 260  & 6 6 6 6 12 & 0.37565 & 0.375679 \\ 
3 & 260  & 6 8 6 8 14 & 0.37570 & 0.375720 \\
3 & 260  & 8 8 6 8 14 & 0.37573 &  0.375721 \\
\end{tabular}
\end{center}
\end{table}

\newpage
\begin{table}
\begin{center}
\caption{\label{MgO421} MgO at a lattice constant of 4.21 \AA. We compare
numerical and analytical derivatives when moving the oxygen ion parallel
to the x-direction. Basis sets of the size $[3s2p]$ were used.}
\begin{tabular}{ccc} 
displacement of oxygen  & analytical derivative 
 & numerical derivative \\
& (x-component) & (x-component) \\
in \% of lattice constant & $E_h/a_B$ & $E_h/a_B$\\
+1 & -0.005923 & -0.00593\\
+2 & -0.012254 & -0.01226\\
+3 & -0.019385 & -0.01940 \\
\end{tabular}
\end{center}
\end{table}


\newpage
\begin{references}
\bibitem{Pulay} P. Pulay, Mol. Phys. {\bf 17}, 197 (1969).
\bibitem{Bratoz} S. Brato\u{z}, in 
{\it Calcul des fonctions d'onde mol{\'e}culaire},
Colloq. Int. C. N. R. S. {\bf 82}, 287 (1958).
\bibitem{PulayAdv} P. Pulay, Adv. Chem. Phys. {\bf 69}, 241 (1987).
\bibitem{PulayChapter} P. Pulay, in {\it Applications of Electronic Structure
Theory}, edited by H. F. Schaefer III, 153 (Plenum, New York, 1977).
\bibitem{Schlegel}H. B. Schlegel, Adv. Chem. Phys. {\bf 67}, 249 (1987).
\bibitem{Helgaker} T. Helgaker and P. J{\o}rgensen, Adv. in Quantum Chem.
{\bf 19}, 183 (1988)
\bibitem{Teramae} H. Teramae, T. Yamabe, C. Satoko and A. Imamura,
Chem. Phys. Lett. {\bf 101}, 149 (1983); H. Teramae, T. Yamabe and
A. Imamura, J. Chem. Phys. {\bf 81}, 3564 (1984).
\bibitem{Kertesz} M. Kertesz, Chem. Phys. Lett. {\bf 106}, 443 (1984).
\bibitem{Suhai} S. Suhai, Phys. Rev. B {\bf 27}, 3506 (1983); S. Suhai,
Chem. Phys. Lett. {\bf 96}, 619 (1983).
\bibitem{SunBartlett} J.-Q. Sun and R. J. Bartlett, J. Chem. Phys. 
{\bf 104}, 8553 (1996).
\bibitem{HirataIwata} S. Hirata and S. Iwata, J. Chem. Phys. {\bf 109},
4147 (1998).
\bibitem{HirataIwataDFT} S. Hirara and S. Iwata, J. Chem. Phys. {\bf 107},
10075 (1997); S. Hirata, H. Torii and M. Tasumi, Phys. Rev. B {\bf 57}, 11994
(1998); S. Hirata and S. Iwata, J. Chem. Phys. {\bf 108}, 7901 (1998); 
S. Hirata and S. Iwata, J. Phys. Chem. A {\bf 102}, 8426 (1998).
\bibitem{HirataIwata2nd} S. Hirata and S. Iwata, J. Mol. Struct.: THEOCHEM
{\bf 451}, 121 (1998).
\bibitem{Jacquemin} D. Jacquemin, J.-M. Andr{\'e} and B. Champagne,
J. Chem. Phys. {\bf 111}, 5306 (1999); J. Chem. Phys. {\bf 111}, 5324 (1999).
\bibitem{KudinScuseria} K. N. Kudin and G. E. Scuseria, Phys. Rev. B
{\bf 61}, 16440 (2000).
\bibitem{Manual} V. R. Saunders, R. Dovesi, C. Roetti, M. Caus\`a, 
N. M. Harrison, R. Orlando, C. M. Zicovich-Wilson, {\sc crystal 98} User's
Manual, Theoretical Chemistry Group, University of Torino (1998).
\bibitem{Pisani1980} C. Pisani and R. Dovesi, Int. J. Quantum Chem.
{\bf 17}, 501 (1980); R. Dovesi and C. Roetti, Int. J. Quantum Chem.
{\bf 17}, 517 (1980).
\bibitem{CRYSTALbuch} C. Pisani, R. Dovesi, and C. Roetti,
{\em Hartree-Fock Ab Initio Treatment of Crystalline Systems}, edited by 
G. Berthier et al, Lecture Notes in Chemistry Vol. 48 (Springer, Berlin,
1988).
\bibitem{Vicbook} V. R. Saunders, N. M. Harrison, R. Dovesi, C. Roetti,
{\it Electronic Structure Theory: From Molecules to Crystals} (in preparation)
\bibitem{VicNATO} V. R. Saunders, in {\it Methods in Computational Molecular
Physics}, edited by G. H. F. Diercksen and S. Wilson, 1 
(Reidel, Dordrecht, Netherlands, 1984).
\bibitem{Andre} J. M. Andr\'e, L. Gouverneur and G. Leroy, Int. J. Quant.
Chem. {\bf 1}, 427 (1967); Int. J. Quant. Chem. {\bf 1}, 451 (1967);
J. M. Andr\'e, J. Chem. Phys. {\bf 50}, 1536 (1969).
\bibitem{Delhalle} J. Delhalle, J. M. Andr\'e, Ch. Demanet and J. L. Br\'edas,
Chem.  Phys. Lett. {\bf 54}, 186 (1978).
\bibitem{Vic1994} V. R. Saunders, C. Freyria-Fava, R. Dovesi, and C. Roetti,
Comp. Phys. Comm. {\bf 84}, 156 (1994).
\bibitem{Parry} D. E. Parry, Surf. Science {\bf 49}, 433 (1975);
{\bf 54}, 195 (1976) (Erratum).
\bibitem{Ewald} P. P. Ewald, Ann. Phys. (Leipzig) {\bf 64}, 253 (1921).
\bibitem{VicCoulomb} V. R. Saunders, C. Freyria-Fava, R. Dovesi, L. Salasco,
and C. Roetti, Mol. Phys. {\bf 77}, 629 (1992).
\bibitem{McMurchieDavidson} L. E. McMurchie and E. R. Davidson,
J. Comput. Phys. {\bf 26}, 218 (1978).
\bibitem{Boys} S. F. Boys, Proc. Roy. Soc. {\bf A 200}, 542 (1950).
\bibitem{McWeeny} R. McWeeny, Nature {\bf 166}, 21 (1950).
\bibitem{Dovesi1983} R. Dovesi, C. Pisani, C. Roetti, and V. R. Saunders,
Phys. Rev. B {\bf 28}, 5781 (1983).
\bibitem{HelgakerTaylor} T. Helgaker and P. R. Taylor, Theor. Chim. Acta
{\bf 83}, 177 (1992).
\bibitem{Harris} F. E. Harris, in {\it Theoretical Chemistry: Advances
and Perspectives}, Vol. 1, 147 (1975), edited by H. Eyring and D. Henderson, 
Academic Press, New York
\end{references}
\end{document}